\documentclass{moriond}

\def\be{\begin{equation}}
\def\ee{\end{equation}}
\def\bea{\begin{eqnarray}}
\def\eea{\end{eqnarray}}

\usepackage{tikz}
\usetikzlibrary{shapes,arrows,positioning,automata,backgrounds,calc,er,patterns}
\usepackage[compat=1.1.0]{tikz-feynman}

\usepackage{subcaption}
\newcommand{\Photo}{}

\begin{document}
\vspace*{4cm}
\title{Impact of CPV phases on flavour violating $H$ and $Z$ decays }

\author{ E. Pinsard$^{1\,}$\footnote{Speaker}, A. Abada$^{2}$, J. Kriewald$^{3}$, S. Rosauro-Alcaraz$^{2}$ and A. M. Teixeira$^{1}$}

\address{$^{(1)}$ Laboratoire de Physique de Clermont (UMR 6533), CNRS/IN2P3, \\ Univ. Clermont Auvergne, 4 Av. Blaise Pascal, 63178 Aubi\`ere Cedex, France
\\ $^{(2)}$ P\^ole Th\'eorie, Laboratoire de Physique des 2 Infinis Irène Joliot Curie (UMR 9012), \\
CNRS/IN2P3,
15 Rue Georges Clemenceau, 91400 Orsay, France \\
${^{(3)}}$ Jožef Stefan Institut, Jamova Cesta 39, P. O. Box 3000, 1001 Ljubljana, Slovenia
}

\maketitle\abstracts{
Standard Model extensions via heavy neutral leptons lead to modifications in the lepton mixing matrix, including new Dirac and Majorana CP violating phases. 
Here we consider the role of the Majorana fermions and of new CP violating phases in Higgs and $Z$-boson lepton flavour violating decays% $H,Z\to \ell_\alpha^\pm\ell_\beta^\mp$
, as well as in the corresponding CP-asymmetries. 
We confirm that these decays are sensitive to the presence of additional sterile states and show that the new CP violating phases may lead to both destructive and constructive interferences in the decay rates.
Interestingly the $Z\to \mu^\pm\tau^\mp$ rates are within FCC-ee reach, with associated CP-asymmetries that can potentially reach up to 30\%.}

\section{Introduction}\label{sec:Intro}

Neutrino oscillation phenomena call for an extension of the Standard Model (SM). Various New Physics (NP) scenarios have been proposed to account for neutrinos masses and the observed lepton mixing pattern. Among them, one minimal and appealing possibility is a SM extension via heavy Majorana sterile states. 
The presence of such heavy neutral leptons (HNL) opens the door to new mixings and CP violating phases (both Dirac and Majorana).

It has been recently shown that the new leptonic CP violating (CPV) phases arising in SM extensions with 2 heavy Majorana fermions can have a significant impact on charged lepton flavour violating (cLFV) observables and even alter the correlation between cLFV observables (that would be present in the CP conserving case).~\cite{Abada:2021zcm}

In~\cite{Abada:2022asx} we carried out a comprehensive study of the role of the HNL, as well as their associated CPV phases on cLFV neutral boson decays. We also discussed how the presence of the CP violating phases can be probed via the CP asymmetries in $Z$-boson decays. In the following, we highlight the most relevant results of this work.

\section{A minimal HNL extension of the SM}\label{sec:model}

In order to study the effects of the CPV phases on cLFV neutral boson decays, we rely on a minimal SM extension via 2 Majorana sterile fermions. The presence of the new states will lead to an enlarged mixing matrix - instead of the $3\times 3$ Pontecorvo-Maki-Nagakawa-Sakata (PMNS) mixing matrix, the lepton mixings are now encoded in the $5\times 5 $ unitary matrix $ \mathcal{U}$. This matrix can be parametrised through a series of 10 rotations $R_{ij}$ (each of them containing a real angle $\theta_{ij}$ as well as a Dirac phase $\delta_{ij}$), and a diagonal matrix which includes the Majorana phases $\varphi_i$. The left-handed mixings are encoded in the $3\times 3$ upper left block of $\mathcal{U}$; this would-be PMNS, which is no longer unitary, leads to modified charged and neutral (lepton) currents. The relevant interaction Lagrangians for $n_S$ sterile states, in the physical basis, can then be cast as
\begin{eqnarray}\label{eq:lagrangian}
 \mathcal{L}_{W^\pm}&=& -\frac{g_w}{\sqrt{2}} \, W^-_\mu \,
\sum_{\alpha=1}^{3} \sum_{j=1}^{3 + n_S} \mathcal{U}_{\alpha j} \bar \ell_\alpha 
\gamma^\mu P_L \nu_j \, + \, \mathrm{H.c.}\,, \nonumber \\
 \mathcal{L}_{Z^0}^{\nu}&=&-\frac{g_w}{4 \cos \theta_w} \, Z_\mu \,
\sum_{i,j=1}^{3 + n_S} \bar \nu_i \gamma ^\mu \left(
P_L {C}_{ij} - P_R {C}_{ij}^* \right) \nu_j\,, \nonumber \\
\mathcal{L}_{H^0}&=& -\frac{g_w}{4 M_W} \, H  \,
\sum_{i\ne j= 1}^{3 + n_S}    \bar \nu_i\,\left[{C}_{ij}\,\left(
P_L m_i + P_R m_j \right) +{C}_{ij}^\ast\left(
P_R m_i + P_L m_j \right) \right] \nu_j\,,
\end{eqnarray}
in which $\alpha = 1, \dots, 3$ denotes the flavour of the charged leptons and $i, j = 1, \dots, 3+n_S$ correspond to the physical (massive) 
neutrino states; $P_{L,R} = (1 \mp \gamma_5)/2$, $g_w$ is the weak coupling constant, and $\cos \theta_w =  M_W /M_Z$. Moreover, the coefficients ${C}_{ij} $ are defined as: ${C}_{ij} = \sum_{\alpha = 1}^3
  \mathcal{U}_{i\alpha}^\dagger \,\mathcal{U}_{\alpha j}^{\phantom{\dagger}}\:$. 

\section{Impact of leptonic CP phases on cLFV neutral boson decays }\label{sec:results}
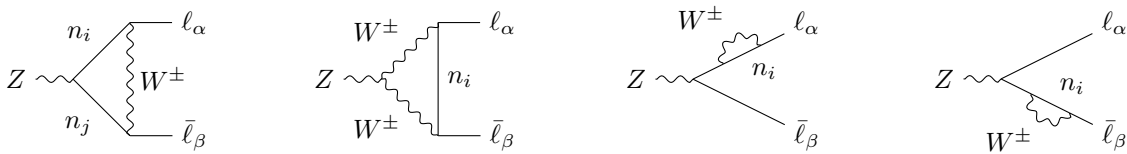
\begin{figure}[h!]
    \centering
    \begin{subfigure}[b]{0.24\textwidth}
    \centering
\begin{tikzpicture}
    {\small\begin{feynman}
    \vertex (a) at (0,0) {\(Z\)};
    \vertex (b) at (0.75,0);
    \vertex (c) at (1.5,0.75);
    \vertex (d) at (1.5,-0.75);
    \vertex (e) at (2.35,0.75) {\( \ell_\alpha\)};
    \vertex (f) at (2.35,-0.75) {\( \bar\ell_\beta\)};
    \diagram* {
    (a) -- [boson] (b),
    (b) -- [ edge label=\(n_i\)] (c),
    (c) -- [boson, edge label=\( W^\pm\)] (d),
    (d) -- [edge label=\(n_j\)] (b),
    (c) -- [] (e),
    (f) -- [] (d)
    };
    \end{feynman}}
    \end{tikzpicture}
    \end{subfigure}
    \hfill
    \begin{subfigure}[b]{0.24\textwidth}
    \centering
       \begin{tikzpicture}
   {\small \begin{feynman}
    \vertex (a) at (0,0) {\(Z\)};
    \vertex (b) at (0.75,0);
    \vertex (c) at (1.5,0.75);
    \vertex (d) at (1.5,-0.75);
    \vertex (e) at (2.35,0.75) {\( \ell_\alpha\)};
    \vertex (f) at (2.35,-0.75) {\( \bar\ell_\beta\)};
    \diagram* {
    (a) -- [boson] (b),
    (b) -- [boson, edge label=\( W^\pm\)] (c),
    (c) -- [ edge label=\(n_i\)] (d),
    (d) -- [boson, edge label=\( W^\pm\)] (b),
    (c) -- [] (e),
    (f) -- [] (d)
    };
    \end{feynman}}
    \end{tikzpicture}
    \end{subfigure}
    \hfill
    \begin{subfigure}[b]{0.24\textwidth}
    \centering
    \begin{tikzpicture}
    {\small\begin{feynman}
    \vertex (a) at (0,0) {\(Z\)};
    \vertex (b) at (0.75,0);
    \vertex (c) at (1.15, 0.2);
    \vertex (d) at (1.65, 0.45);
    \vertex (e) at (2.25, 0.75) {\( \ell_\alpha\)};
    \vertex (f) at (2.25,-0.75) {\( \bar\ell_\beta\)};
    
    \diagram* {
    (a) -- [boson] (b),
    (b) -- [] (c) -- [ edge label'=\(n_i\)] (d) -- [] (e),
    (b) -- [] (f),
    (c) -- [boson, half left, edge label=\( W^\pm\)] (d)
    };   
    \end{feynman}}
    \end{tikzpicture}
    \end{subfigure}
    \hfill
    \begin{subfigure}[b]{0.24\textwidth}
    \centering
    {\small\begin{tikzpicture}
     \begin{feynman}
    \vertex (a) at (0,0) {\(Z\)};
    \vertex (b) at (0.75,0);
    \vertex (c) at (1.15, -0.2);
    \vertex (d) at (1.65,-0.45);
    \vertex (e) at (2.25, 0.75) {\( \ell_\alpha\)};
    \vertex (f) at (2.25,-0.75) {\( \bar\ell_\beta\)};
    
    \diagram* {
    (a) -- [boson] (b),
    (b) -- [] (c) -- [edge label=\(n_i\)] (d) -- [] (f),
    (b) -- [] (e),
    (c) -- [boson, half right, edge label'=\( W^\pm\)] (d)
    };   
    \end{feynman}
    \end{tikzpicture}}
    \end{subfigure}
    \caption{Feynman diagrams contributing to cLFV $Z$ decays (in unitary gauge).}
    \label{fig:cLFVZdecays:UG}
\end{figure}

The modified currents allow for tree-level flavour changing interactions and will also contribute to cLFV boson decays through one-loop vertex corrections and charged lepton self-energy diagrams, as it can be seen in Fig.~\ref{fig:cLFVZdecays:UG} (notice that Higgs decays receive analogous contributions).

\begin{figure}
\centering{\includegraphics[width=0.5\linewidth]{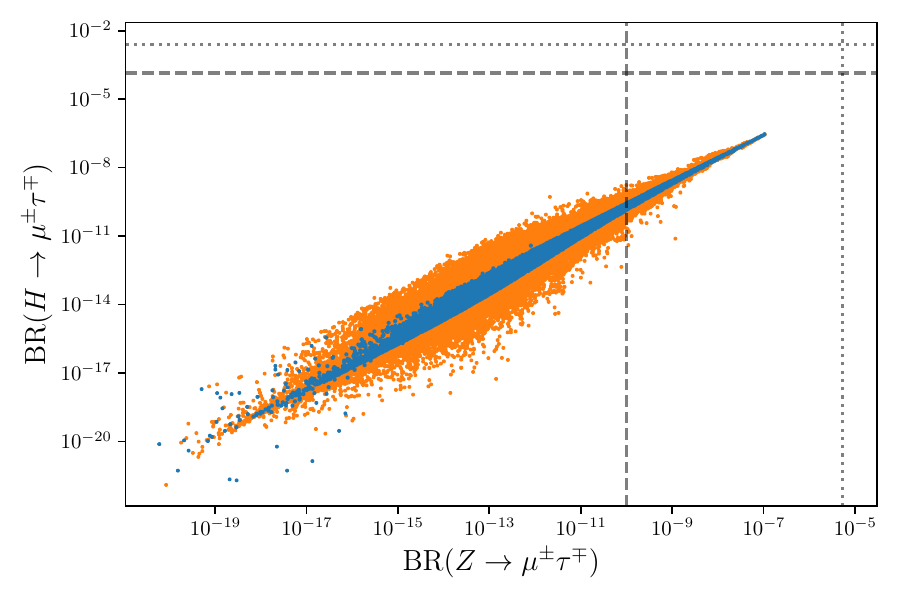}}
\caption[]{BR($H \to \mu \tau$) and BR($Z \to \mu \tau$) in the ``3+2 model'' parameter space. All active-sterile mixing angles and Dirac and Majorana CP phases are randomly varied. Blue points correspond to vanishing phases, while orange denotes random values of all phases ($\delta_{\alpha i}$ and $\varphi_i$, with $\alpha = e,\mu,\tau$ and $i=4,5$). Dotted (dashed) lines denote current bounds (future sensitivity).}
\label{fig:higgs_z}
\end{figure}

Both Higgs and $Z$ decays share a common topology of the one-loop contributions, leading to analogous behaviours of these two observables. Their decay rates~\footnote{We carried out the computation without any approximation and the form factors are given in.~\cite{Abada:2022asx} Our predictions for the cLFV $Z$ decay rate agree with previous results obtained in the limit of vanishing final state lepton masses.~\cite{DeRomeri:2016gum,Illana:2000ic}} receive the dominant contribution from the diagram with two virtual neutrinos in the loop, and thus are significantly impacted by the HNL masses - increasing with the heaviest HNL mass. In what follows we focus on TeV range HNL masses (with $m_4=5$~TeV and $m_5-m_4 \in [10~\mathrm{MeV}, 1~\mathrm{TeV}]$), since one is led to cLFV $Z$ decays within future FCC-ee sensitivity~\cite{FCC:2018byv} while complying with all relevant experimental constraints.
Moreover we will only discuss the $\mu-\tau$ sector~\footnote{The decays into $e\mu$ and $e\tau$ final states are discussed in.~\cite{Abada:2022asx}} since the branching ratio for $Z\to\mu\tau$ is larger than for the other cLFV channels, thus offering more promising experimental prospects. 
A comparison between $H\to \mu\tau$ and $Z\to \mu\tau$ is shown in Fig.~\ref{fig:higgs_z}, in which one can see, as anticipated, that both observables are strongly correlated in the CP conserving case. This behaviour is however affected by the presence of the new CPV phases that induce interference effects, both constructive and destructive. Notice that $H\to \mu\tau$ is far beyond future experimental reach, while $Z\to \mu\tau$ could be observable at a future FCC-ee (running at the $Z$-pole mass), as shown by the dashed lines. 

\section{CP asymmetries in cLFV boson decays}\label{sec:asymmetries}

In view of the capabilities of a future FCC-ee, we consider the potential contributions of this minimal model to CP-asymmetries in $Z$-boson decay, $ \mathcal{A}_{CP}(Z\to\ell_\alpha\ell_\beta)$, defined as 
\begin{equation}\label{eq:def:ACP:Z}
    \mathcal{A}_{CP}(Z\to\ell_\alpha\ell_\beta)\, =\, 
    \frac{
    \Gamma(Z \to \ell_\alpha^-\ell_\beta^+) -
    \Gamma(Z \to \ell_\alpha^+\ell_\beta^-)
    }{\Gamma(Z \to \ell_\alpha^-\ell_\beta^+) +
    \Gamma(Z \to \ell_\alpha^+\ell_\beta^-)}\,.
\end{equation}
Again the dominant contributions to the asymmetries stem from the loop diagram containing two HNL. The Dirac CPV phases, especially $\delta_{34}$, have a significant impact on $\mathcal{A}_{CP}(Z\to\ell_\alpha\ell_\beta)$, while Majorana phases have a reduced impact. However the latter will lead to more significant contributions provided that Dirac phases are also present. 

\begin{figure}
\begin{minipage}{0.5\linewidth}
\centerline{\includegraphics[width=0.95\linewidth]{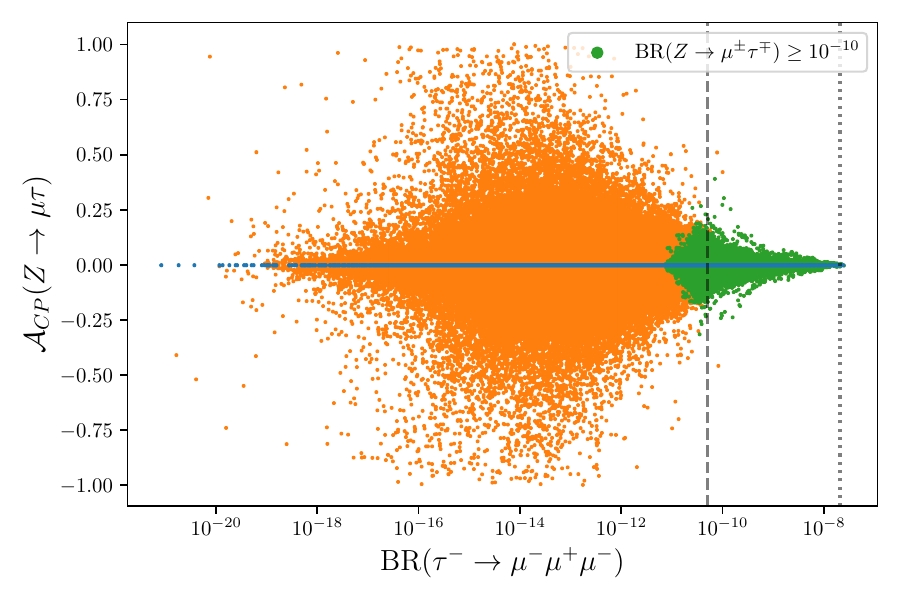}}
\end{minipage}
\hfill
\begin{minipage}{0.5\linewidth}
\centerline{\includegraphics[width=0.95\linewidth]{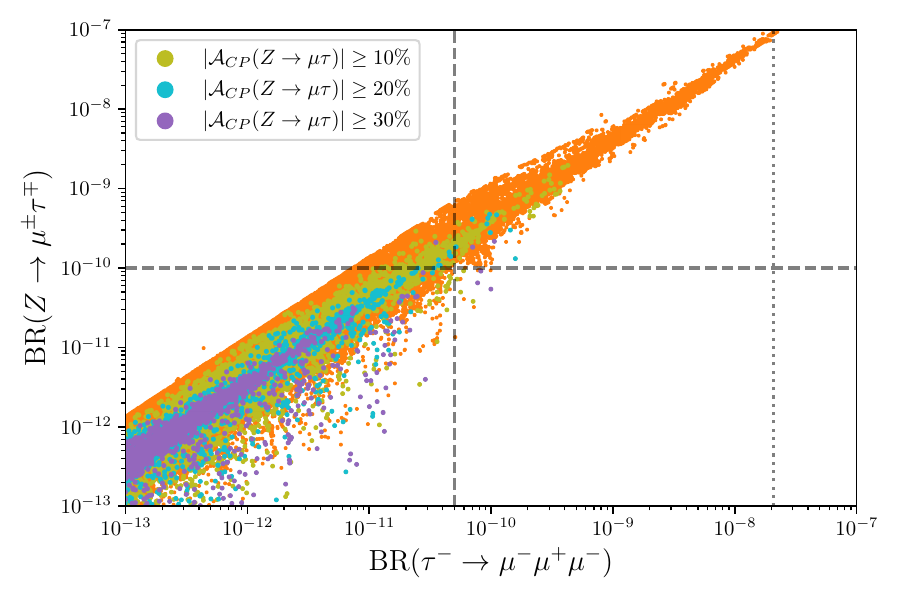}}
\end{minipage}
\caption[]{On the left, prospects for $\mathcal{A}_{CP} (Z \to \mu \tau)$ vs.~BR($\tau\to \mu\mu\mu$): orange and blue points as indicated in Fig.~\ref{fig:higgs_z}; in green, points associated with $\mathrm{BR}(Z\to\mu^\pm\tau^\mp) \geq 10^{-10}$, within future reach of FCC-ee. On the  right, $\mathrm{BR}(Z\to\mu^\pm\tau^\mp)$ vs.~$\mathrm{BR}(\tau\to\mu\mu\mu)$. Colour code as before, with olive green, cyan and purple points respectively denoting $|\mathcal A_{CP}(Z\to\mu\tau)|\geq 10\%\,, 20\%\, \mathrm{ and } \,30\%$.}
\label{fig:z_ACP}
\end{figure}

Complying with all relevant experimental constraints, the CP-asymmetries in $Z\to\mu\tau$ can be very large (up to 100\%), as it can be seen in the left panel of Fig.~\ref{fig:z_ACP}, where we present the prospects for $\mathcal{A}_{CP}(Z\to\mu\tau)$ vs. the branching fraction of $\tau \to 3\mu$. For $\mathrm{BR}(\tau\to 3\mu)$ and $\mathrm{BR}(Z \to \mu\tau)$ both within future sensitivity of FCC-ee, the CP-asymmetries can still reach up to 20\%. In the right panel of Fig.~\ref{fig:z_ACP}, we display a complementary view of these results, showing that in the $\mu-\tau$ sector one can simultaneously test the presence of HNL as well as their CP violating phases via several observables. The joint observation of these three observables would be highly suggestive of such a SM extension via \textit{at least} 2 heavy Majorana states.

\medskip 
Finally, in order to further emphasise the role of the CP-asymmetries, we have selected the following CP conserving ($P_1$) and CP violating ($P_2$) benchmark points: 
\begin{eqnarray}
    P_1 &:& m_4 = 5 \:\mathrm{TeV}, \,
    m_5 = 5.1 %5141
    \:\mathrm{TeV},\nonumber \\
    &\phantom{:}&
    s_{14} = -0.0028,\, s_{15} = 0.0045,\, s_{24} = -0.0052,\, s_{25} = -0.0037,\, s_{34} = -0.052,\, s_{35} = -0.028,
    \nonumber \\
    &\phantom{:}& \delta_{ij} = \varphi_{i} =0\,, \nonumber\\
    P_2 &:& m_4 = 5 \:\mathrm{TeV}, \, m_5 =5.1     %5084
    \:\mathrm{TeV},     \nonumber \\
    &\phantom{:}&
    s_{14} = 0.0002,\, s_{15} = -7.1\times 10^{-5},\, s_{24} = -0.0024,\, s_{25} = 0.029,\, s_{34} = -0.073,\, s_{35} = -0.037,\nonumber\\
    &\phantom{:}& \delta_{14} =0.71,\, \delta_{15} = 5.21,\, \delta_{24} = 2.06,\, \delta_{25} = 4.78,\, \delta_{34} = 3.80,\, \delta_{35} = 4.74, %\nonumber \\
    %&\phantom{:}&
        \varphi_{4} = 1.77\,,\, \varphi_5 = 4.33\,.\nonumber
\end{eqnarray}
$P_1$ and $P_2$ lead to similar cLFV predictions both lying within future sensitivity,~\footnote{With $\mathrm{BR}(\mu\to 3e) = 2\times 10^{-15}$, $\mathrm{CR}(\mu-e \, ,\mathrm{Al}) = 5\times10^{-14}$, $\mathrm{BR}(\tau\to 3\mu) = 1\times 10^{-10}$, $\mathrm{BR}(Z\to\mu\tau) = 2\times 10^{-10} $.} and thus rendering them indistinguishable if any cLFV signal would be observed in the future. It is worth noticing that $P_2$ features smaller mixing angles than $P_1$; the identical predictions for the cLFV observables is due to the role of CPV phases, which are at the source of constructive interferences. From a phenomenological point of view, these two regimes are indistinguishable, hence little can be learnt about the presence of CPV phases. However, $P_2$ leads to non vanishing CP-asymmetries with $\mathcal{A}_{CP}(Z\to\mu\tau) = 30\,\%$, offering a clear distinction between CP conserving and CP violating scenarios.

\section{Conclusions}\label{sec:conclusions}

Heavy neutral lepton extensions of the SM introduce new leptonic CPV phases that can have a significant impact on cLFV observables and their interpretation. We have illustrated these results within a minimal BSM construction, in which 2 HNL are added to the SM content. In particular, the presence of the new TeV scale states (and their CPV phases) leads to $\mathrm{BR}(Z\to\mu\tau)$ within future FCC-ee sensitivity and the associated CP-asymmetries, that can reach up to 30\%, is a key ingredient to disentangle between CP conserving and CP violating regimes.

\section*{Acknowledgments}

This project has received support from the European Union's Horizon 2020 research and innovation programme under the Marie Sk\l{}odowska-Curie grant agreement No.~860881 (HIDDe$\nu$ network) and from the IN2P3 (CNRS) Master Project, ``Flavour probes: lepton sector and beyond'' (16-PH-169).

\end{document}